\documentclass{optica-article}

\journal{opticajournal} 

\articletype{Research Article}

\usepackage{lineno}

\begin{document}

\title{On the localisation of the high-intensity region of simultaneous space-time foci}

\author{Emily Archer,\authormark{1,*} Bangshan Sun,\authormark{2}, Roman Walczak\authormark{1,3}, Martin Booth\authormark{2} and Simon Hooker\authormark{1}}

\address{\authormark{1}Department of Physics, University of Oxford, Clarendon Laboratory, Parks Rd, Oxford, OX1 3PU, UK\\
\authormark{2}Department of Engineering Science, University of Oxford, Parks Rd, Oxford, OX1 3PJ, UK\\
\authormark{3}Somerville College, Woodstock Rd, Oxford, OX2 6HD, UK}

\email{\authormark{*}emily.archer@desy.de Present address: Deutsches Elektronen-Synchrotron DESY, Notkestraße 85, 22607 Hamburg} 


\begin{abstract*} 
Simultaneous space-time focusing (SSTF) is sometimes claimed to reduce the longitudinal extent of the high-intensity region near the focus, in contradiction to the original work on this topic. Here we seek to address this confusion by using numerical and analytical methods to investigate the degree of localisation of the spatio-temporal intensity of an SSTF pulse. The analytical method is found to be in excellent agreement with numerical calculations and yields, for bi-Gaussian input pulses, expressions for the three-dimensional spatio-temporal intensity profile of the SSTF pulse, and for the on-axis bandwidth, pulse duration, and pulse-front tilt (PFT) of the SSTF pulse. To provide further insight, we propose a method for determining the transverse input profile of a non-SSTF pulse with equivalent spatial focusing. We find that the longitudinal variations of the peak axial intensities of the SSTF and spatially equivalent (SE) pulses are the same, apart from a constant factor, and hence that SSTF does not constrain the region of high intensity more than a non-SSTF pulse with equivalent focusing. We demonstrate that a simplistic method for calculating the pulse intensity exaggerates the degree of intensity localisation, unless the spatio-temporal couplings inherent to SSTF pulses are accounted for.
\end{abstract*}

\section{Introduction}
Simultaneous space-time focusing (SSTF) is a phenomenon that has been exploited for well over a decade in the fields of microscopy and laser machining \cite{Zhu:05,Durst:06,j.optcom.2007.05.071,lanier_nonlinear_2016,sun_four-dimensional_2018}. It localises time-dependent processes to a particular axial depth by `focusing' the pulse in time to achieve full temporal recompression (shortest pulse duration) only at the focus. The benefits of SSTF in applications such as multiphoton imaging \cite{Zhu:05,Durst:06,j.optcom.2007.05.071,sun_four-dimensional_2018} and laser machining \cite{lanier_nonlinear_2016,sun_four-dimensional_2018} have been discussed by several authors.

\begin{figure}[tb]
	\begin{center}
		\includegraphics[width=\linewidth]{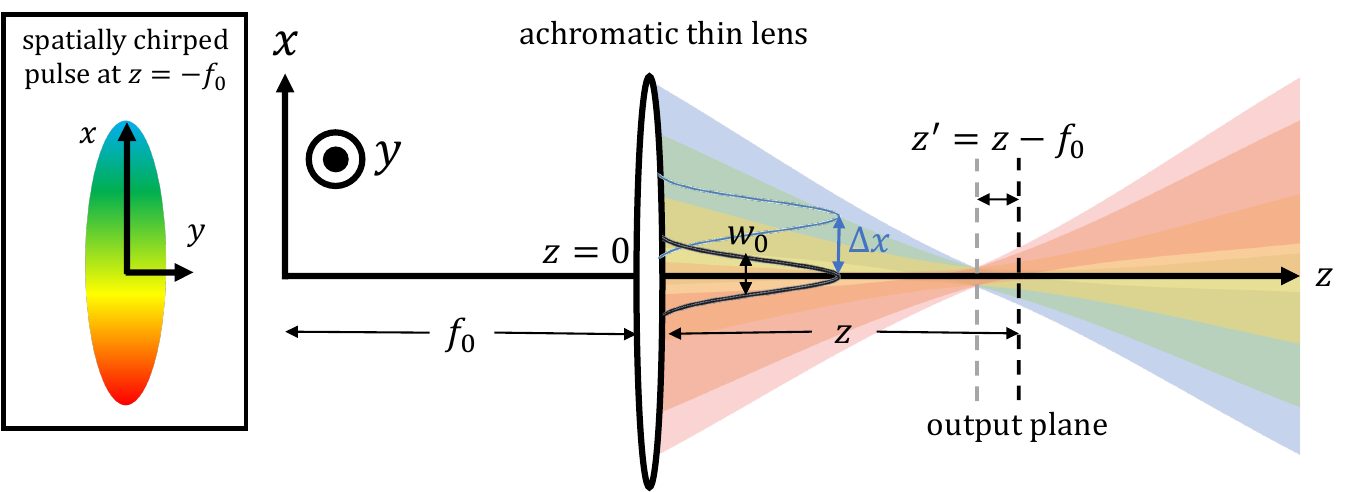}
		\caption{Schematic of a linear simultaneous space-time focus. A pulse chirped transversely along the $x$-axis is focused by a lens and propagates a distance $z$ to the output plane. The nominal Gaussian beamlet is shown for the central frequency $\omega_0$ in black and for an arbitrary beamlet with a different frequency $\omega$ in blue, where $w_0$ is the spot size and $\Delta x$ is the transverse distance between the two different frequencies.}
		\label{fig:simultaneousspacetime} 
	\end{center}
 \vspace{-10pt}
\end{figure}

As illustrated in Figure~\ref{fig:simultaneousspacetime}, simultaneous space-time focusing is achieved by introducing a spatial chirp --- i.e.\ displacing transversely each frequency within the pulse by a different amount --- before focusing the spatially-chirped pulse with an achromatic optic, such as a lens. SSTF is sometimes claimed to lead to tighter localisation of the region of high intensity near the focus. The standard argument proposed for this is as follows: as the pulse focuses spatially, the increase in local bandwidth --- i.e.\ the bandwidth at a particular spatial location --- decreases the pulse duration, which increases the intensity more rapidly than would occur from the reduction in transverse size of the pulse alone. However, introducing a spatial chirp also increases the transverse size of the \emph{input} pulse. This, on its own, reduces the focal spot size of the beam in the direction of the spatial chirp, and reduces the Rayleigh range of the beam in that dimension. Therefore, to understand how SSTF works it is important to disentangle the effects of tighter focusing from those arising from the changes in local bandwidth.

There appears to be some confusion in the literature about these points. In the original papers on this topic \cite{Zhu:05,Durst:06,j.optcom.2007.05.071} it is \emph{not} stated that SSTF localizes the region of high intensity more tightly, only that it can localise the region in which a time-dependent process occurs, due to the reduction in pulse duration. For example, Durst et al. \cite{Durst:06} state that ``the axial characteristics of SSTF [in two-photon excitation fluorescence (TPEF)] are determined by the input spatial profile''. This contrasts with later work \cite{LI20172898,OE.20.014244}, which refers to the ``axial localisation of intensity with SSTF''. If SSTF does indeed lead to greater localisation of the high-intensity region, then it could be used to localise purely intensity-dependent processes, such as the optical field ionisation of gases \cite{PhysRevLett.62.1259}. Possible applications would then include ionisation injection in laser-wakefield accelerators \cite{HiddingCathode}.

The aim of this paper is to clarify whether SSTF does or does not improve the degree of axial localisation of the high intensity region near the focus. We define, for a given SSTF pulse, a method for determining the spatially equivalent non-SSTF pulse and use numerical methods, and an analytical approximation, to demonstrate that SSTF does \emph{not} localize the region of high intensity any more than a non-SSTF pulse with spatially equivalent focusing. The analytical approach yields, for the case of a bi-Gaussian input pulse which has a Gaussian transverse spatial profile and a Gaussian spectrum, analytical expressions for the three-dimensional spatio-temporal intensity profile of the SSTF pulse, and for the axial evolution of the peak intensity, pulse duration, bandwidth, temporal chirp, and the pulse-front tilt (PFT). Finally, we show that a commonly used method for calculating the pulse intensity fails for SSTF pulses if it does not account for the pulse-front tilt, and that this can lead to an overestimate of the degree of intensity localisation.  

\begin{figure}[tb]
    \centering
    \includegraphics[width=.8\linewidth]{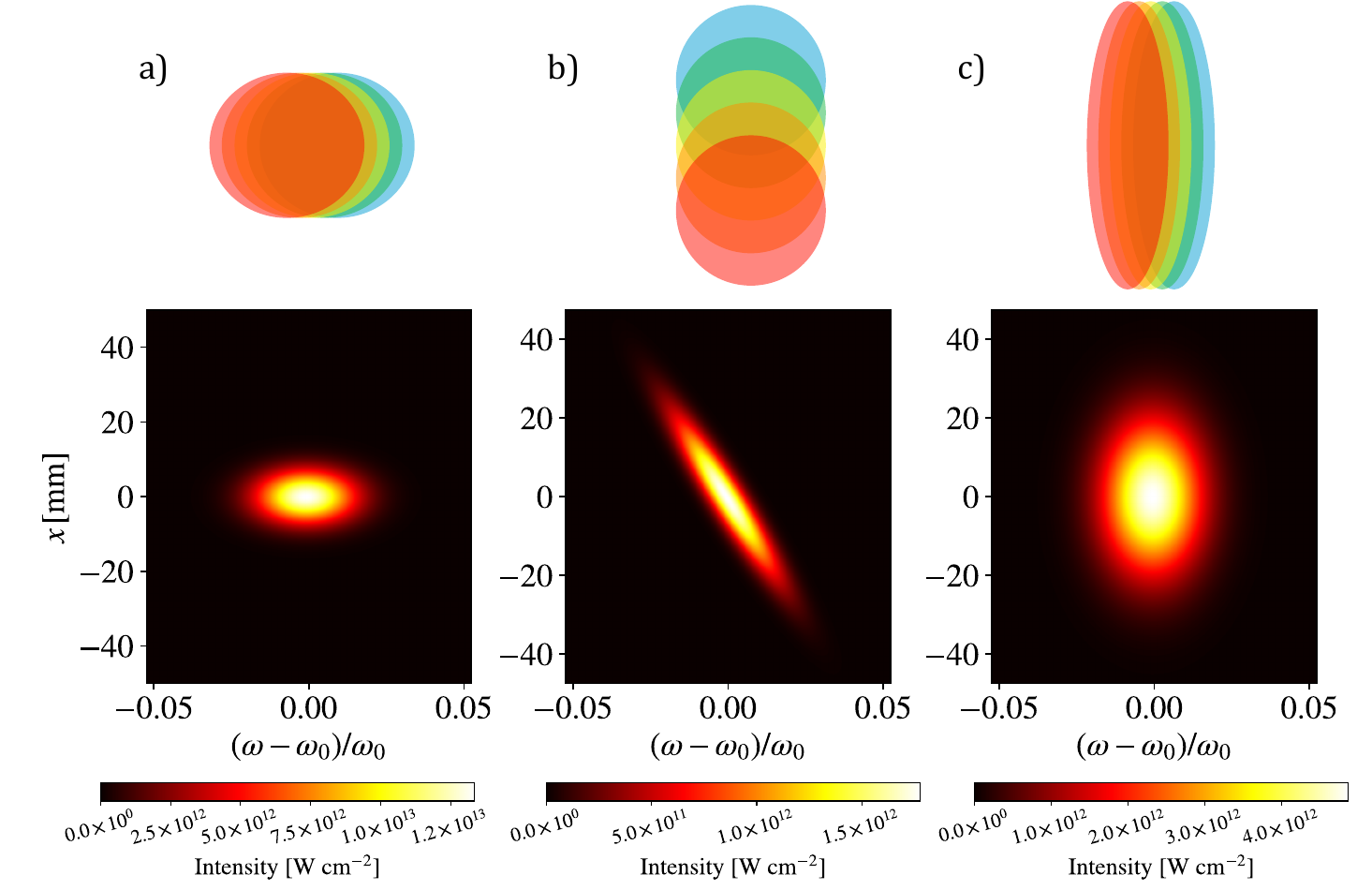}
    \caption{Spatio-spectral intensity profiles in the input plane ($z = -f_0$) before being focused by the lens for: (a) the polychromatic beam ($\alpha=0$), prior to introduction of the spatial chirp; (b) the SSTF pulse ($\alpha=-3$), after introduction of spatial chirp in the $x$-direction; and (c) the spatially-equivalent pulse ($\alpha=-3$) with same spatial amplitude as the SSTF but with all frequency components overlapped spatially.}
    \label{fig:faircomparison}
\end{figure}

\section{Model of SSTF}
SSTF can be modelled by considering the input pulse as a superposition of monochromatic Gaussian beamlets which are offset from the optical axis by a transverse distance $\Delta x(\omega) = \alpha(\omega-\omega_0)$, where $\omega$ and $\omega_0$ are respectively the angular frequencies of the beamlet and the centre of the spectrum of the input pulse. The chirp scaling factor $\alpha$ is used to control the direction and strength of the spatial chirp and all stated values of $\alpha$ are normalised by the factor $\alpha_{\textrm{norm}}=w_0/\Delta\omega$, where $\Delta\omega$ is the full-width at half maximum (FWHM) bandwidth of the input pulse.

We assume that each beamlet propagates in the $z$-direction, and that it has a waist of spot size $w_0$ (defined as the radius at which the intensity is $1/\mathrm{e}^2$ of the peak value) located in the front focal plane of a lens of focal length $f_0$. The lens is taken to be located at the origin (see the supplemental information for a diagram of the coordinate system). 

In the input plane ($z = -f_0$) the amplitude of each beamlet may then be written as,

\begin{align}
    \tilde{U}^\mathrm{in}(x,y,-f_0,\omega) &= \frac{- i z_\mathrm{R} \tilde{\mathcal{U}}(\omega)}{q(-f_0)} \exp \left[\frac{i k}{2 q(-f_0)}\left\{\left[x - \Delta x(\omega)\right]^2 + y^2\right\}\right],\label{Eq:InputLinearSSTF}
\end{align}
where $\Delta x(\omega) = \alpha(\omega-\omega_0)$ and generally, the complex beam parameter $q(z) = z - z_0 - i z_\mathrm{R}$, such that $q(-f_0) = - i z_\mathrm{R}$. Here $z_0=-f_0$ is the position of the input beam waist and $z_\mathrm{R} = \pi w_0^2/\lambda$ is the Rayleigh range, where $w_0$ is the spot size at the waist, and $\lambda$ is the wavelength.

We assume that the pulse has a Gaussian spectrum $\tilde{\mathcal{U}}(\omega) = (2\Gamma)^{-1/2}\exp{[-(\omega-\omega_0)^2/(4\Gamma)]}$, peaked at the central frequency $\omega_0$, where $\Gamma = a(1 + i\chi)$ is a complex parameter that characterises the temporal width and chirp of the pulse, with $a=2\ln{2}/\tau_{\textrm{p}0}^2$, in which $\tau_{\textrm{p}0}$ is the FWHM fully-compressed pulse duration of the intensity profile. For simplicity, in this work we set the input temporal chirp factor $\chi=0$.\par

\subsection{Spatially-equivalent pulse}\label{Sec:FairComparison}
As discussed above, for an SSTF pulse, the degree of intensity localisation around the focus could depend on the increased input beam size in the direction of the spatial chirp, as well as on the longitudinal variation of the pulse duration. In order to disentangle these effects we define a `spatially equivalent' (SE) pulse $\tilde{U}_\mathrm{SE}$ that has the same transverse amplitude profile as the SSTF pulse, but no spatio-temporal coupling. We define the input transverse amplitude profile of the SE pulse to be equal to the integral of the modulus of the input SSTF field over the spectrum of the pulse, in other words, the transverse profile of the SE pulse is taken to be the projected amplitude of the SSTF pulse.  As described in the supplemental document, the full spatio-spectral amplitude of the SE pulse is then given by: (i) multiplying by the spectrum of the SE pulse; and (ii) multiplying by a constant factor that ensures that the SSTF and SE pulses have the same total energy. The input SE pulse is therefore defined by:

\begin{align}
    \tilde{U}_\mathrm{SE}^\mathrm{in}(x,y,-f_0,\omega) &= \left(1 + \frac{4\alpha^2\Gamma}{w_0^2}\right)^{-1/4} \frac{\tilde{\mathcal{U}}(\omega) }{\sqrt{q_x(-f_0) q(-f_0)}} \exp \left[\frac{i k x^2}{2 q_x(-f_0)} + \frac{i k y^2}{2 q(-f_0)}\right],\label{Eq:FairComparison}
\end{align}
where,

\begin{align}
    w_{0x}&= w_0 \sqrt{1 + \frac{4\alpha^2\Gamma}{w_0^2}}.\label{Eq:w0x}
\end{align}
Figure~\ref{fig:faircomparison} shows, schematically and quantitatively, the spatio-spectral intensity profiles of the input SSTF and SE pulses. The spatial chirp of the SSTF pulse is evident in Figure~\ref{fig:faircomparison}(b); in contrast, the SE pulse shown in in Figure~\ref{fig:faircomparison}(c) is seen to have the same transverse extent as the SSTF pulse, but with no coupling between the transverse position and the spectrum. Table \ref{tab:parametersSSTF} gives the laser parameters assumed throughout this work.

\begin{table}[!b]
    \centering
    \begin{tabular}{r|c}
         Pulse energy, $\mathcal{E}$ & $\SI{1}{\J}$ \\
         Central wavelength, $\lambda_0$ & $\SI{800}{\nm}$ \\
         Fully-compressed pulse duration, $\tau_{\textrm{p}0}$ & $\SI{45}{\fs}$ \\
         Input spot size, $w_0$ & $\SI{10}{\mm}$ \\
         Focal length, $f_0$ & $\SI{1}{\m}$ \\
    \end{tabular}
    \caption{Table of laser parameters for SSTF simulations.}
    \label{tab:parametersSSTF}
\end{table}
\begin{figure}[!tb]
    \centering
    \includegraphics[width=.6\linewidth]{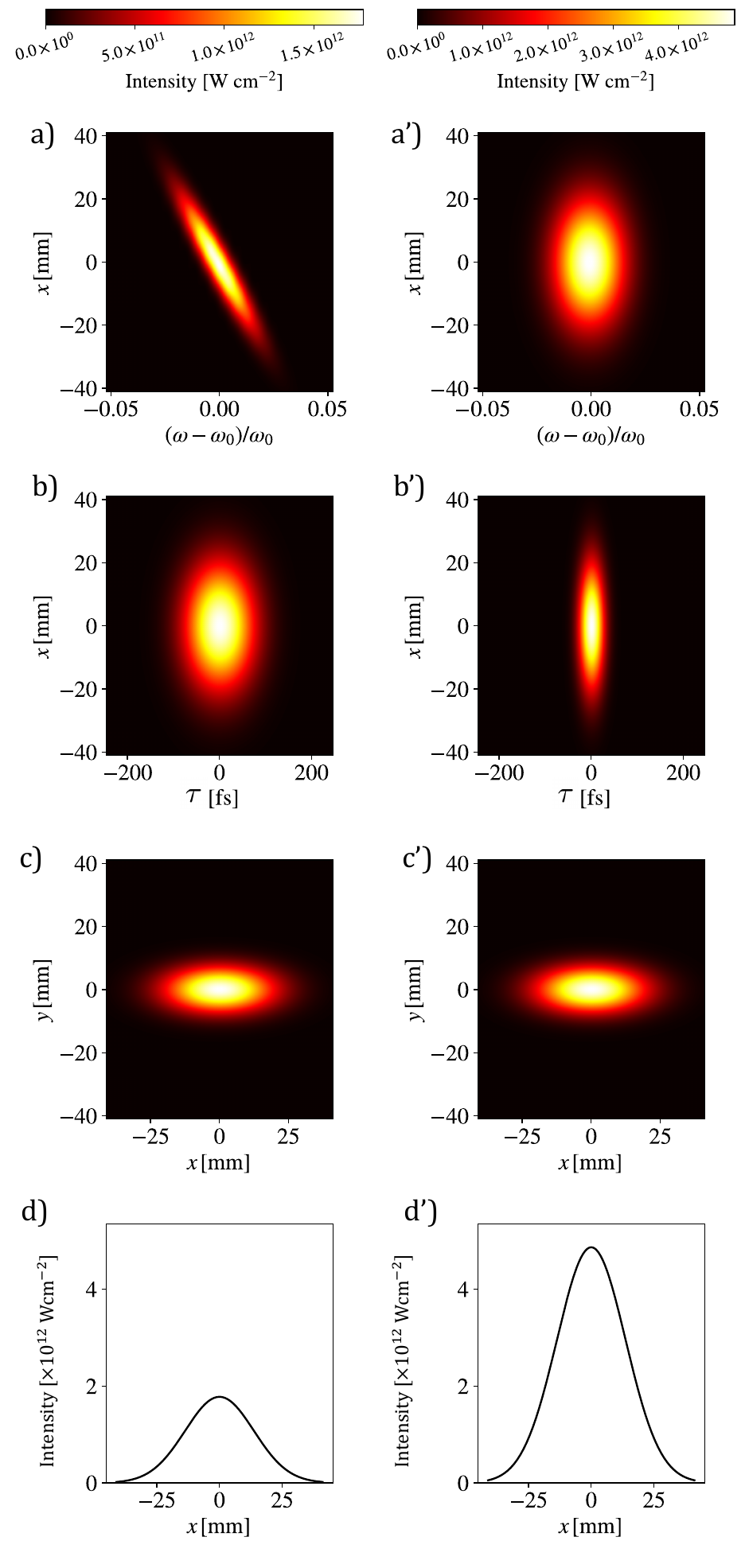}
    \caption{Comparison between the intensity distributions of the input SSTF pulse (left column) and the corresponding SE pulse (right column) before being focused by the lens for the case $\alpha = -3$. The pulse parameters are given in Table~\ref{tab:parametersSSTF}.  (a), (a') Show the spatio-spectral domain for $y=0$; (b), (b') show the spatio-temporal domain for $y=0$; (c), (c') show the transverse spatial profile at the temporal peak of the laser pulse $\tau=0$; and (d), (d') are lineouts of the data in (c), (c') at $y=0$.}
    \label{fig:inputlinear}
\end{figure}

Figure~\ref{fig:inputlinear} provides further details of the spatio-spectral and spatio-temporal intensity profiles, in the input plane, for both the SSTF and SE pulses. The correlation between frequency and transverse position for the SSTF pulse can clearly be seen in Figure~\ref{fig:inputlinear}(a), whereas this is absent for the SE pulse. Figs~\ref{fig:inputlinear}(b) and (b$^{\prime}$) show that, for both SSTF and SE input pulses, the temporal intensity profiles are independent of the transverse coordinate $x$. However, it can be seen that the FWHM duration of the SSTF pulse is longer (\SI{123}{\fs}) than that of the SE pulse (\SI{45}{\fs}), owing to the reduced local bandwidth of the SSTF pulse. The global bandwidth --- i.e.\ the spatially integrated bandwidth --- always remains the same because the pulse propagates in vacuum and is assumed to be the same for both the SSTF and SE pulses. Figs~\ref{fig:inputlinear}(c) and (c$^{\prime}$) show that the transverse intensity profiles of the SSTF and SE pulses are identical in shape, which is a consequence of the definition of the SE pulse. However, as shown in Figures~\ref{fig:inputlinear}(d) and (d$^{\prime}$), the SE pulse has a greater on-axis intensity owing to its shorter pulse duration.

\subsection{Numerical treatment}\label{Sec:Numerical_treatment}
The field beyond the lens can be calculated numerically by the Collins method \cite{Collins:70}, which gives the spatio-spectral amplitude as, 

\begin{align}
	\tilde{U}^{\textrm{out}}(\mathbf{r},\omega) = \frac{1}{i\lambda B} e^{ik(z-z_0)}\int_{-\infty}^{\infty}\int_{-\infty}^{\infty}\tilde{U}^{\textrm{in}}(\mathbf{r},\omega) e^{ikS}dx_0dy_0,\label{Eq:OutputFieldCollins}\\
	S = \bigg\{\frac{1}{2B}\Big[A(x_0^2+y_0^2)+D(x^2+y^2)-2(xx_0+yy_0)\Big]\bigg\}\label{Eq:Collins},
\end{align}
where $A$, $B$, $C$, and $D$ are the elements of the transfer matrix describing the propagation of a ray from the input plane $z=-f_0$ to the output plane $z=z$. The integral can be evaluated numerically for arbitrary input amplitude $\tilde{U}^{\textrm{in}}(\mathbf{r}_0,\omega)$. Here, however, we assume the spatio-spectral amplitudes given by Eqs.\,\ref{Eq:InputLinearSSTF} and  \ref{Eq:FairComparison} for the SSTF and SE pulses respectively. The ray transfer matrix is calculated in the supplemental document. The Rayleigh range after the lens $z'_\mathrm{R} = f_0^2/z_\mathrm{R}$ is used to normalise the longitudinal axis. In performing the calculations, a hybrid grid is used which increases the axial resolution from $\SI{764}{\micro\meter}$ to $\SI{306}{\micro\meter}$ for the region within $z'_\mathrm{R}$ of the focus. The Collins method gives $\tilde{U}^\mathrm{out}(x,y,\omega)$ at a particular axial position $z$, which can then be numerically Fourier transformed to the time-domain to give $U^\mathrm{out}(x,y,t)$ at each axial position $z$; repeating this for different longitudinal axial positions $z$ in a grid that spans a number of Rayleigh ranges around the focus gives the full spatio-temporal profile $U^\mathrm{out}(x,y,z,t)$. The fields are then shifted from the lab-frame coordinate $t$ to the light-speed coordinate $\tau$, as is explained in the supplemental document.

\subsection{Analytical treatment}\label{Sec:Analytic_treatment}
As described in detail on the supplemental document, for an SSTF pulse formed by Gaussian input beamlets defined by Eqs.\,\ref{Eq:InputLinearSSTF} and \ref{Eq:FairComparison}, the spatio-spectral amplitude of the SSTF pulse in the region beyond the lens is given by,

\begin{align}
    \tilde{U}^\mathrm{out}(x,y,z,\omega)  &=\tilde{\mathcal{U}}(\omega) \frac{q'(0)}{q(0)} \frac{- i z_\mathrm{R}}{q'(z)} \exp[i k (f_0 + z)] \exp \left[\frac{ik}{2 q'(z)} \left(x^2 + y^2\right)\right]\notag\\
    &\hspace{80pt}\times \exp \left\{- \frac{1}{w_0^2 q'(z)}\left[\Delta x(\omega)^2 (z- f_0) + 2 f_0 x \Delta x(\omega)\right]\right\},\label{Eq:AnalyticField}
\end{align}
where,

\begin{equation}
    q'(z) = z - z'_0 - iz'_\mathrm{R} = z - f_0 - \frac{2if_0^2}{k w_0^2}.\label{Eq:qparameter}
\end{equation}
is the complex radius for a Gaussian beam beyond the lens, as given by the ABCD law.

Similarly, the spatio-spectral amplitude of the SE pulse in the region beyond the lens is found to be,

\begin{align}
    \tilde{U}_\mathrm{SE}^\mathrm{out}(x,y,z,\omega) 
    &= \tilde{\mathcal{U}}(\omega) \exp \left[i (k z + f_0)\right] f_0 \left(1 + \frac{4\alpha^2\Gamma}{w_0^2}\right)^{-1/4} \notag\\
    &\hspace{80pt}\times\sqrt{\frac{1}{q'_x(z)q'(z)}} \exp \left[ \frac{ik}{2} \left[\frac{x^2}{q'_x(z)} + \frac{y^2}{q'(z)}\right]\right\},
\end{align}
where 
\begin{align}
    q_x'(z) &= z - f_0 - i z'_\mathrm{R,eff},
\end{align}
with an effective Rayleigh range $z'_\mathrm{R,eff}$ given by,

\begin{align}
    z'_\mathrm{R,eff} &= \frac{z'_\mathrm{R}}{1 + \frac{4 \alpha^2}{w_0^2}\Gamma}. \label{Eq:Eff_ZR}
\end{align}

The spatio-temporal profile of the SSTF and SE pulses can also be determined analytically if certain approximations are made. First, the frequency dependence of the complex Gaussian parameters is neglected when taking the Fourier transform of the spatio-spectral field to obtain the spatio-temporal field. Second, it is assumed that $\Gamma$ is real. With these assumptions the spatio-temporal amplitudes of the SSTF and SE pulses in the region beyond the lens can be shown to be:

\begin{align}
    U_\mathrm{env}^\mathrm{out}(x,y,z,\tau) &=U_0 \frac{q'(0)}{q(0)} \sqrt{\frac{\Gamma'(z)}{\Gamma}} \frac{- i z_\mathrm{R}}{q'(z)} \exp \left[\frac{ik_0}{2 q'(z)} \left(x^2 + y^2\right)\right] \exp\left[ - \Gamma'(z) \left( \tau - i \frac{2 \alpha f_0 x}{w_0^2 q'(z)}\right)^2\right]\\
    U_\mathrm{SE, env}^\mathrm{out}(x,y,z,\tau) &= U_0 f_0 \left(1 + \frac{4\alpha^2\Gamma}{w_0^2}\right)^{-1/4}
\sqrt{\frac{1}{q'_x(z)q'(z)}} \exp \left[\frac{i k_0}{2} \left[\frac{x^2}{q'_x(z)} + \frac{y^2}{q'(z)}\right]\right\} \exp \left(- \Gamma \tau^2 \right).
\end{align}
where,

\begin{align}
    \frac{1}{\Gamma'(z)} &= \frac{1}{\Gamma} + \frac{4 \alpha^2 }{w_0^2} \frac{z'}{q'(z)}.\label{Eq:Gamma_prime}
\end{align}
and the retarded time is given by,

\begin{align}
    \tau &= t - \frac{f_0 + z}{v_\mathrm{g}},
\end{align}
in which $v_\mathrm{g}$ is the group velocity of the medium surrounding the lens.

\section{Results}\label{Sec:LinearResults}

\begin{figure}[!tb]
    \centering
    \includegraphics[width=.6\linewidth]{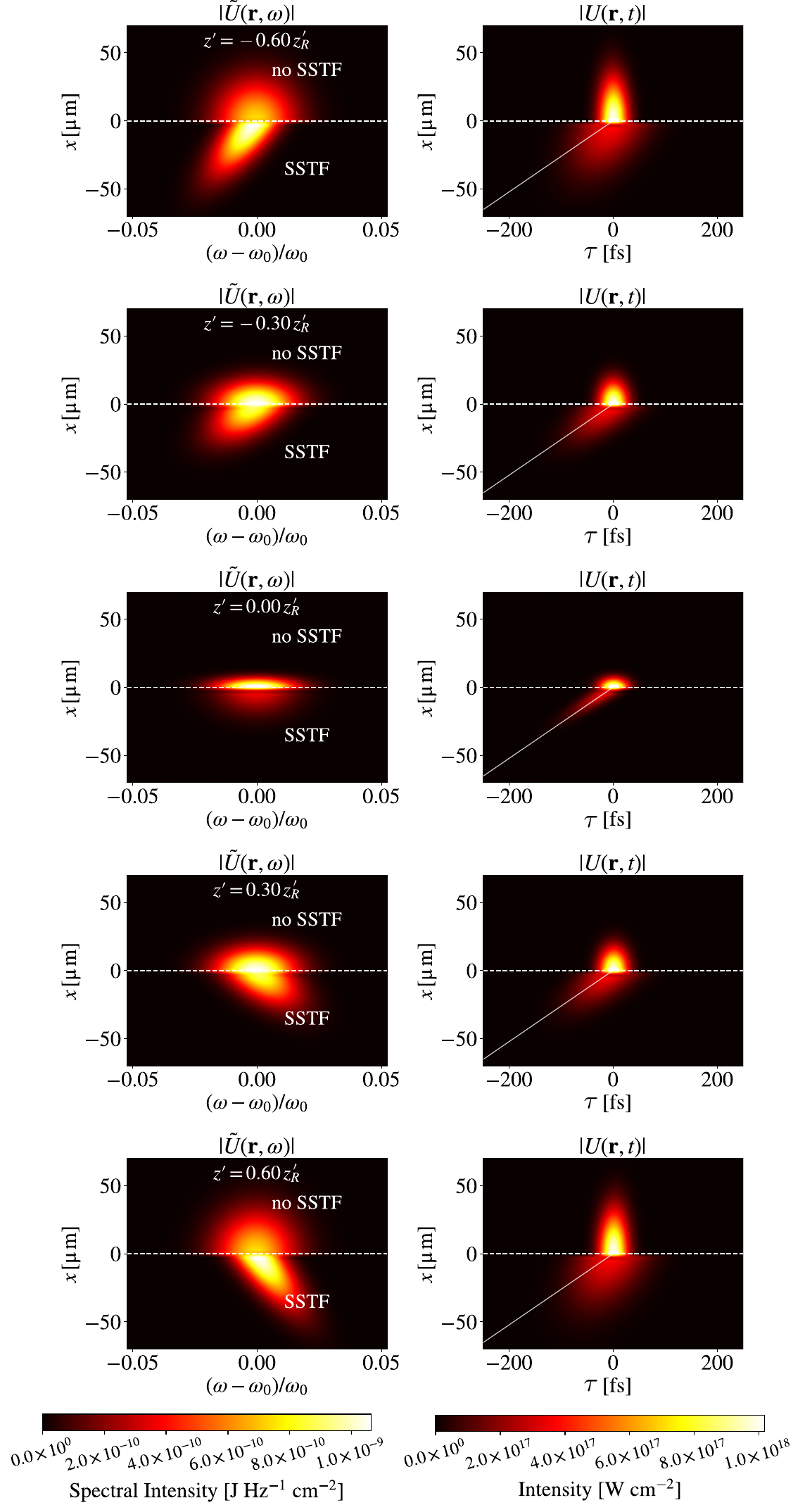}
    \caption{Evolution of the spatio-spectral (left column) and spatio-temporal (right column) intensity profiles of the SSTF and SE pulses (for $\alpha=-3$) in the focal region. The lower half of each plot shows the intensity profile of the SSTF pulse, and the upper half shows that for the SE pulse, both plotted for $y=0$. The $z^{\prime}$ positions at which the fields are calculated are given for each panel in the left column. The solid white line shows the PFT predicted by Eq.\,\ref{Eq:PFTevo}.}
    \label{fig:linearevolution}
\end{figure}

In this section, we consider the evolution of some key features and parameters of the SSTF pulse, before studying the degree of axial localisation. Numerical calculation of the spatio-spectral fields was performed using the Collins method described in Section \ref{Sec:Numerical_treatment}, and results were compared with key analytical results derived in the supplementary document.

Figure~\ref{fig:linearevolution} shows the spatio-spectral (left column) and spatio-temporal (right column) intensity distributions at various axial positions $z'=z-f_0$ relative to the focus $f_0$. The most striking feature of Figure~\ref{fig:linearevolution} is that the transverse chirp of the SSTF pulse away from focus is converted to a pulse-front tilt (PFT) \cite{Akturk:05} --- i.e.\ the arrival time of the peak of the pulse depends linearly on transverse position --- as previously observed for an SSTF pulse \cite{Coughlan:09}. This effect is absent for the SE pulse, as expected. The analytical model gives the angle $\theta(z)$ of the PFT, measured from the $x$-axis, as,

\begin{align}
    \tan \theta (z) &= - \frac{\alpha \omega_0}{f_0} \frac{1}{1 + \left(\frac{z'}{\sqrt{z'_\mathrm{R}z'_\mathrm{R,eff}}} \right)^2}.\label{Eq:PFTevo}
\end{align}
The results shown in Figure~\ref{fig:linearevolution} are in excellent agreement with these analytical expressions. We note that it has previously been shown that the PFT cannot be compensated by applying a counter-acting PFT to the input pulse \cite{Zhang:14}.

\begin{figure}[hbt!]
    \centering
    \includegraphics[width=\linewidth]{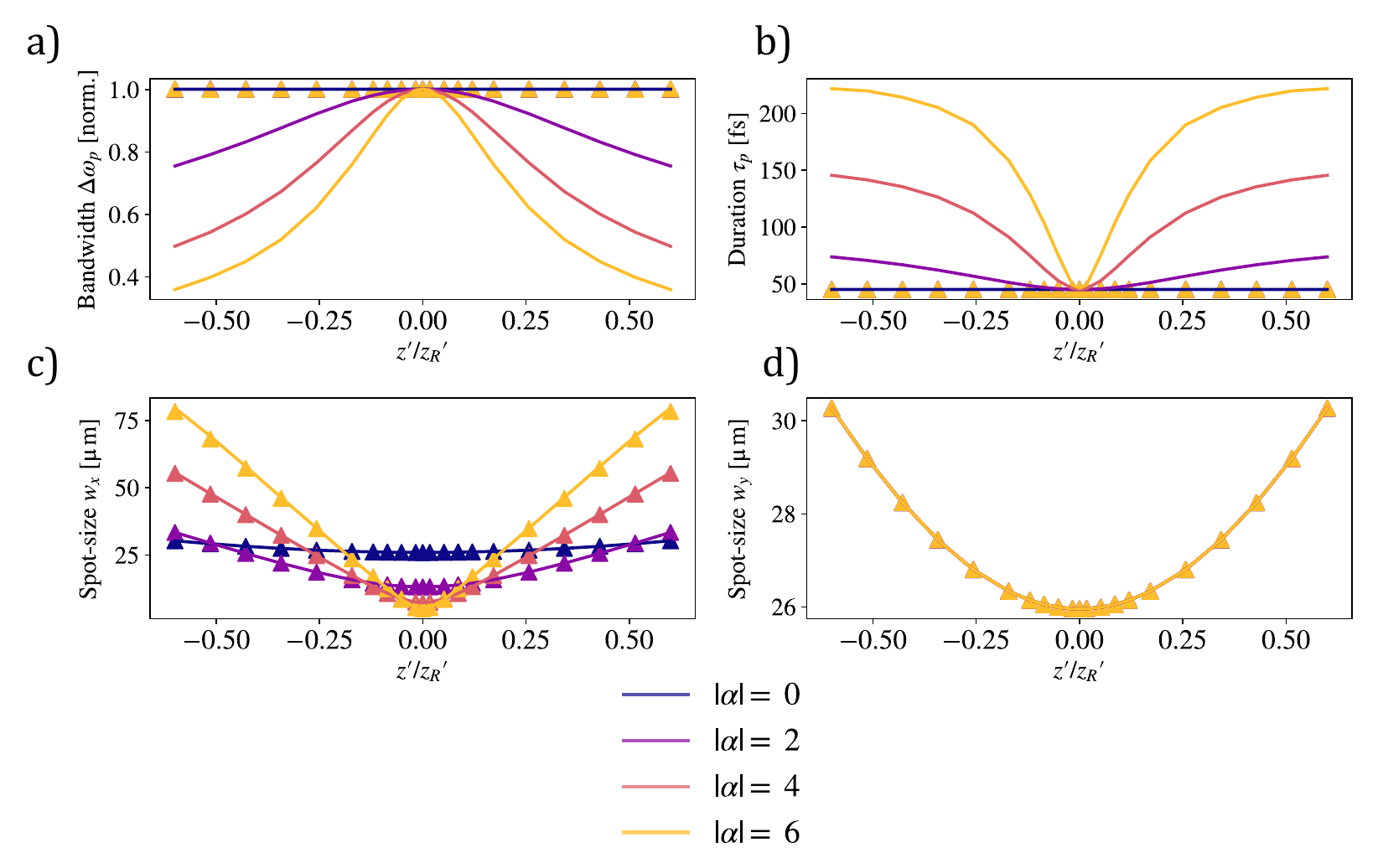}
    \caption{Variation with $z^{\prime}$ of key on-axis parameters of the numerically simulated SSTF and SE pulses, showing: (a) the FWHM bandwidth $\Delta\omega_\mathrm{p}(z)$; (b) the FWHM pulse duration $\tau_{\textrm{p}}(z)$; (c) the spot size $w_x(z)$ in the $x$-direction; and (d) the spot size $w_y(z)$ in the $y$-direction . For all plots, the SSTF and SE pulses are indicated by the solid lines and triangles respectively.}
    \label{fig:outputlinearparameters}
\end{figure}

\subsection{Centroid, Ellipsoidal Spot Size, Bandwidth and Pulse Duration}\label{Sec:KeyParameter}
In order to understand the evolution of the peak axial intensity of the SSTF pulse in the focal region, we first determine the longitudinal dependence of the key parameters of the pulse from the numerical simulations as follows.

For each longitudinal position $z$, the time at which the peak axial intensity occurs is found. At this time, a cross-section of the pulse in the $xy$-plane is taken. Along the $x$-axis, the $\textrm{D}4\Omega$ beam diameter is calculated in terms of the second moment of the transverse intensity distribution $I(x,y)$\cite{ISO11146,Siegman:98}. The $1/\mathrm{e}^2$ Gaussian spot size $w_x$ can then be calculated from the $\textrm{D}4\Omega_x$ beam diameter as $w_x = \textrm{D}4\Omega_{\textrm{x}}/2$. The same method is used to calculate the spot size $w_y$. The on-axis full-width at half maximum (FWHM) bandwidths and pulse durations are calculated for each value of $z$ by recording the intensity profile that passes that point on the optical axis in the spectral and temporal domains, respectively.

The evolution of these key parameters is plotted in Figures~\ref{fig:outputlinearparameters}(a)--(d). Figures~\ref{fig:outputlinearparameters}(a) and (b) confirm that, for the SSTF pulse, the on-axis bandwidth increases (and the pulse duration decreases) as the focus is approached. Away from the focus, the on-axis pulse duration can increase from the fully-compressed value of $\SI{45}{\fs}$ by as much as a factor of $4.5$ for the highest chirp considered ($\alpha=\pm 6$). Also evident in Figure~\ref{fig:outputlinearparameters}(c) is the decrease of the focal spot size in the $x$-direction with increasing  $|\alpha|$. As expected, no such change in focal spot size is observed in the $y$-direction.

\begin{figure}[!tb]
    \centering
    \includegraphics[width=.6\linewidth]{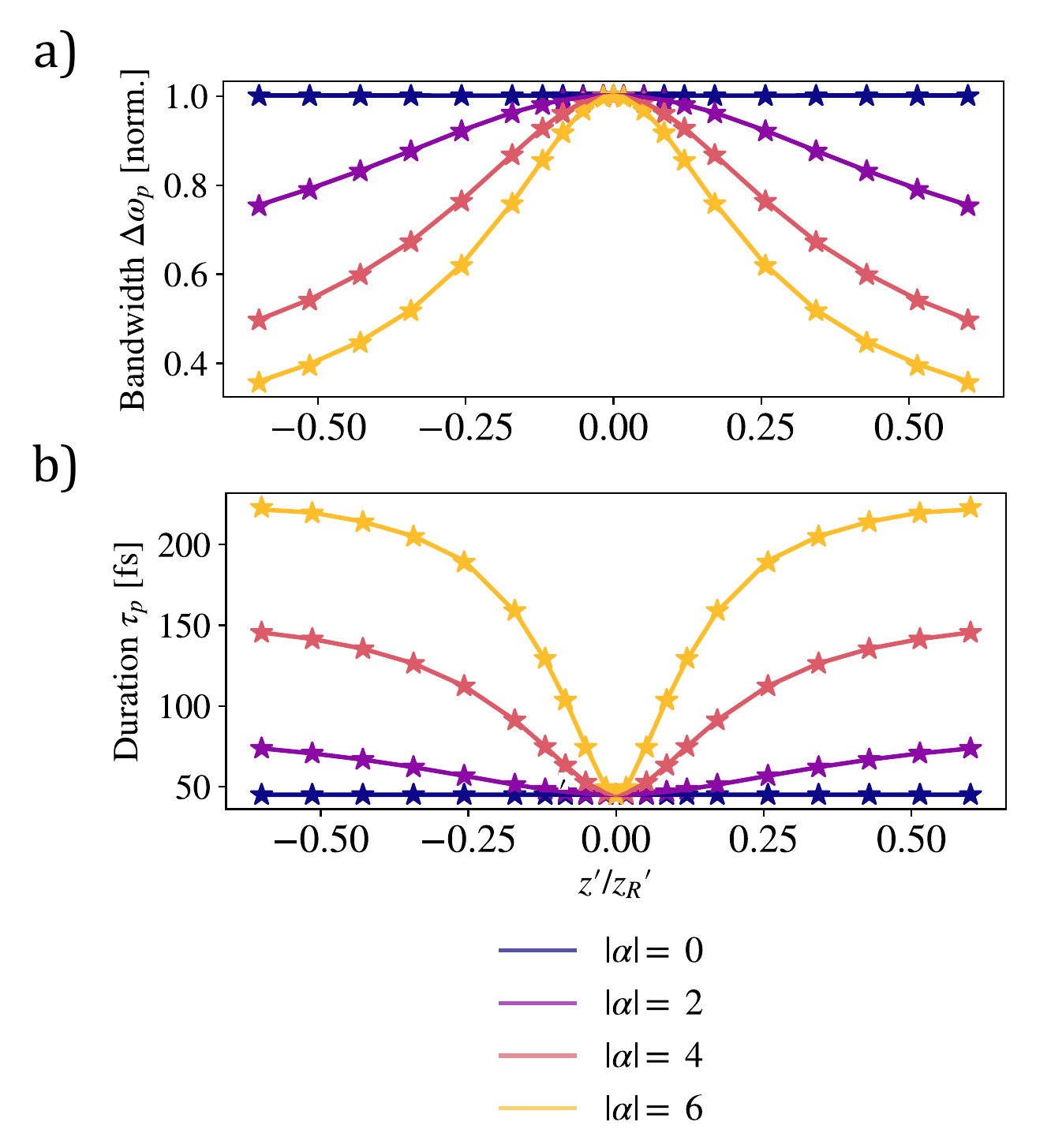}
    \caption{Comparison of (a) the on-axis FWHM bandwidth calculated directly from the fields (solid lines) with the analytical expression given by Eq.\,\ref{Eq:Bandwidth} (stars); (b) the on-axis FWHM pulse duration calculated directly from the fields (solid lines) with the analytical expression given by Eq.\,\ref{Eq:Duration} (stars).}
    \label{fig:parametersanalytic}
\end{figure}

The analytical model gives the on-axis FWHM bandwidth and duration of the SSTF pulse as, 

\begin{align}
    \Delta \omega_\mathrm{p}(z) &= \Delta \omega_{\mathrm{p}0}\sqrt{\frac{1 + \left(\frac{z'}{z'_\mathrm{R}}\right)^2}{1 + \left(\frac{z'}{\sqrt{z'_\mathrm{R}z'_\mathrm{R,eff}}}\right)^2}},\label{Eq:Bandwidth}
\end{align}
and,

\begin{align}
    \tau_\mathrm{p}(z) = \tau_{\mathrm{p}0} \sqrt{\frac{1+\left(\frac{z'}{z'_\mathrm{R,eff}} \right)^2}{1 + \left(\frac{z'}{\sqrt{z'_\mathrm{R}z'_\mathrm{R,eff}}}\right)^2}},\label{Eq:Duration}
\end{align}
respectively, where the effective Rayleigh range $z'_\mathrm{R,eff}$ is given by Eq.\,\ref{Eq:Eff_ZR}. Figure~\ref{fig:parametersanalytic} compares these results with the numerical results shown in Figure~\ref{fig:outputlinearparameters} for different values of $|\alpha|$. Excellent agreement is obtained. As expected, the results are the same for positive and negative values of $\alpha$.

\begin{figure}[!tb]
    \centering
    \includegraphics[width=.6\linewidth]{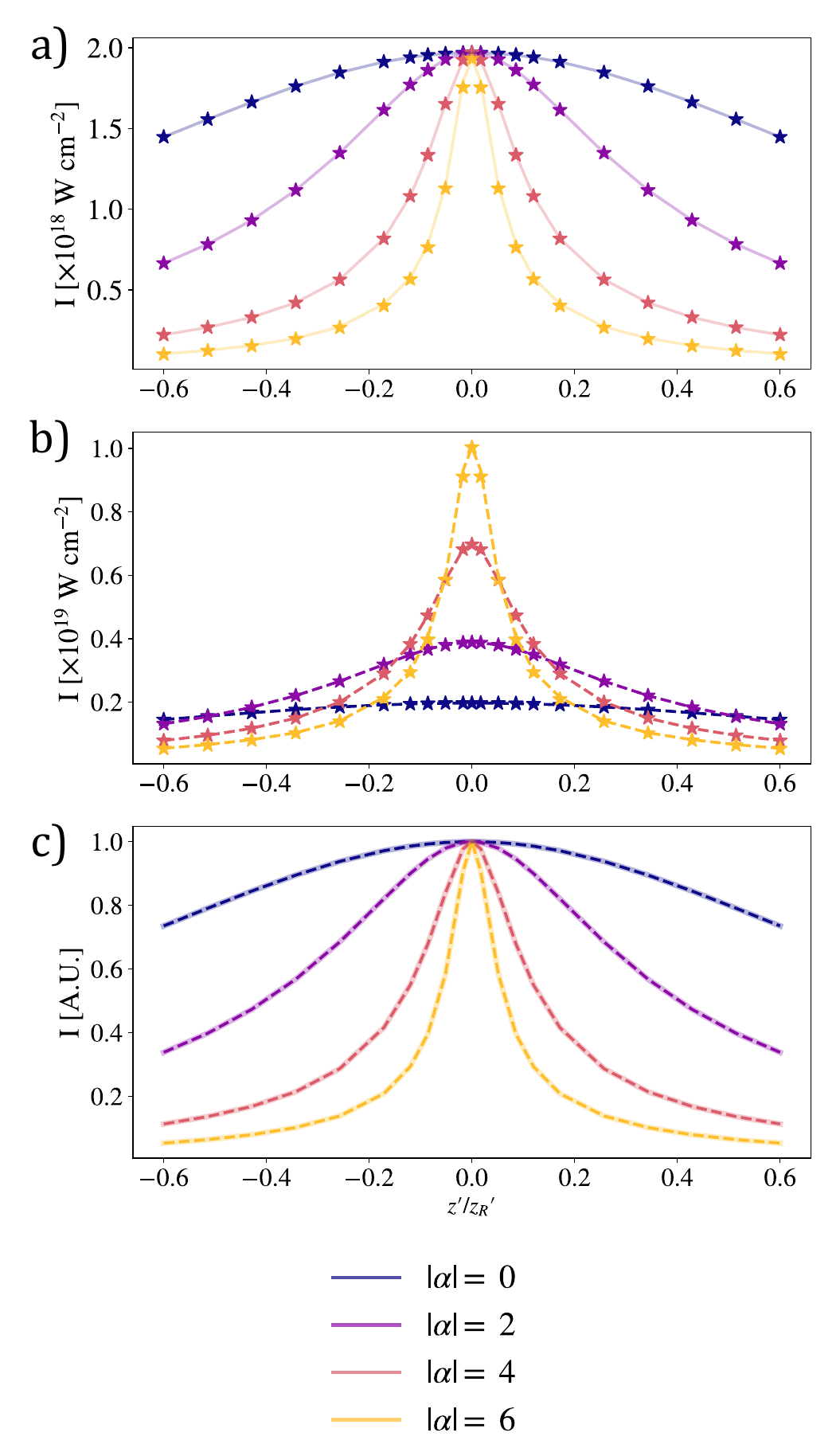}
    \caption{Longitudinal variation of the peak axial intensity calculated numerically from the fields as $I_{\textrm{peak}}=\varepsilon_0 c |\mathbf{E}|_{\textrm{max}}^2$ (lines) with the analytical results (stars) for: (a) the SSTF pulse, Eq.\,\ref{Eq:AnalyticInt}; and (b) the SE pulse, Eq.\,\ref{Eq:AnalyticIntFC}. In (c) the longitudinal variation of the \emph{normalised} peak axial intensities of the SSTF (solid lines) and SE (dashed lines) pulses are compared.}
    \label{fig:intensitypeak}
\end{figure}

\subsection{Variation of peak axial intensity}\label{Sec:AxialConfinement}
Having established how the key parameters of the SSTF pulse evolve in the focal region, we now compare the longitudinal variations of the peak intensities of the SSTF and SE pulses. Figure~\ref{fig:intensitypeak}(a) shows the variation with $z'$ of the peak axial intensity $I_{\textrm{peak}}(z)=\max{\{\varepsilon_0 c |U^\mathrm{out}(x,y,z,t)|^2}\}$ of the SSTF pulse. For convenience, in this work we assume an input pulse energy of $\mathcal{E}=\SI{1}{J}$. We see that as the magnitude of the spatial chirp increases, the region of high intensity becomes more tightly localised around the focus. The degree of localisation is independent of the sign of the spatial chirp, which makes sense since reversing the sign of the spatial chirp does not change the initial transverse distance of each frequency component from the optical axis. A further striking feature of Figure~\ref{fig:intensitypeak}(a) is that the peak intensity of the SSTF pulse at focus ($z'=0$) is independent of $\alpha$. 

These findings are in agreement with the analytical model, which gives the peak axial intensity of the SSTF pulse as,

\begin{align}
     I^\mathrm{out}_\mathrm{peak}(z) &= \left(\frac{z_\mathrm{R} U_0}{f_0}\right)^2\frac{1}{\sqrt{1 + \left(\frac{z'}{z'_\mathrm{R}}\right)^2} \sqrt{1 + \left(\frac{z'}{z'_\mathrm{R,eff}}\right)^2}}.\label{Eq:AnalyticInt}
\end{align}
This result shows that the peak axial intensity is independent of $\alpha$ and that the region of high intensity becomes more localised as $|\alpha|$ increases, since $z'_\mathrm{R,eff}$ decreases with $|\alpha|$. Figure~\ref{fig:intensitypeak}(a) compares directly the longitudinal evolution of the peak axial intensity of the SSTF pulse calculated from Eq.\,\ref{Eq:AnalyticInt} (stars) with the numerical results (solid lines). Excellent agreement is observed.

Figure~\ref{fig:intensitypeak}(b) shows the longitudinal variation of the peak axial intensity for the SE pulse. It can be seen that, unlike for the SSTF pulse, the peak axial intensity at focus increases with $|\alpha|$. This finding is in agreement with the analytical model, for which the peak axial intensity of the SE pulse is found to be,

\begin{align}
    I^\mathrm{SE}_\mathrm{peak}(z) &= \left(\frac{z_\mathrm{R} U_0}{f_0}\right)^2  \sqrt{1 + \frac{4\alpha^2\Gamma}{w_0^2}}\frac{1}{\sqrt{1 + \left(\frac{z'}{z'_\mathrm{R}}\right)^2} \sqrt{1 + \left(\frac{z'}{z'_\mathrm{R,eff}}\right)^2}}.\label{Eq:AnalyticIntFC}
\end{align}
It can be seen that the analytical model predicts that \emph{the longitudinal variations of the peak axial intensities of the SSTF and SE pulses are identical.} Further, on axis the peak intensity of the SE pulse is everywhere larger than that of the SSTF pulse by a constant factor that increases with $|\alpha|$. These results are consistent with the numerical results shown in Fig~\ref{fig:intensitypeak}(c), which shows excellent agreement between the normalized longitudinal profiles of the peak axial intensities of the SSTF and SE pulses. Figure~\ref{fig:intensitypeak}(b) compares directly the longitudinal evolution of the peak axial intensity of the SE pulse calculated from Eq.\,\ref{Eq:AnalyticIntFC} (stars) with the numerical results (dashed lines). Again, excellent agreement is observed.

\begin{figure}[!tb]
    \centering
    \includegraphics[width=.6\linewidth]{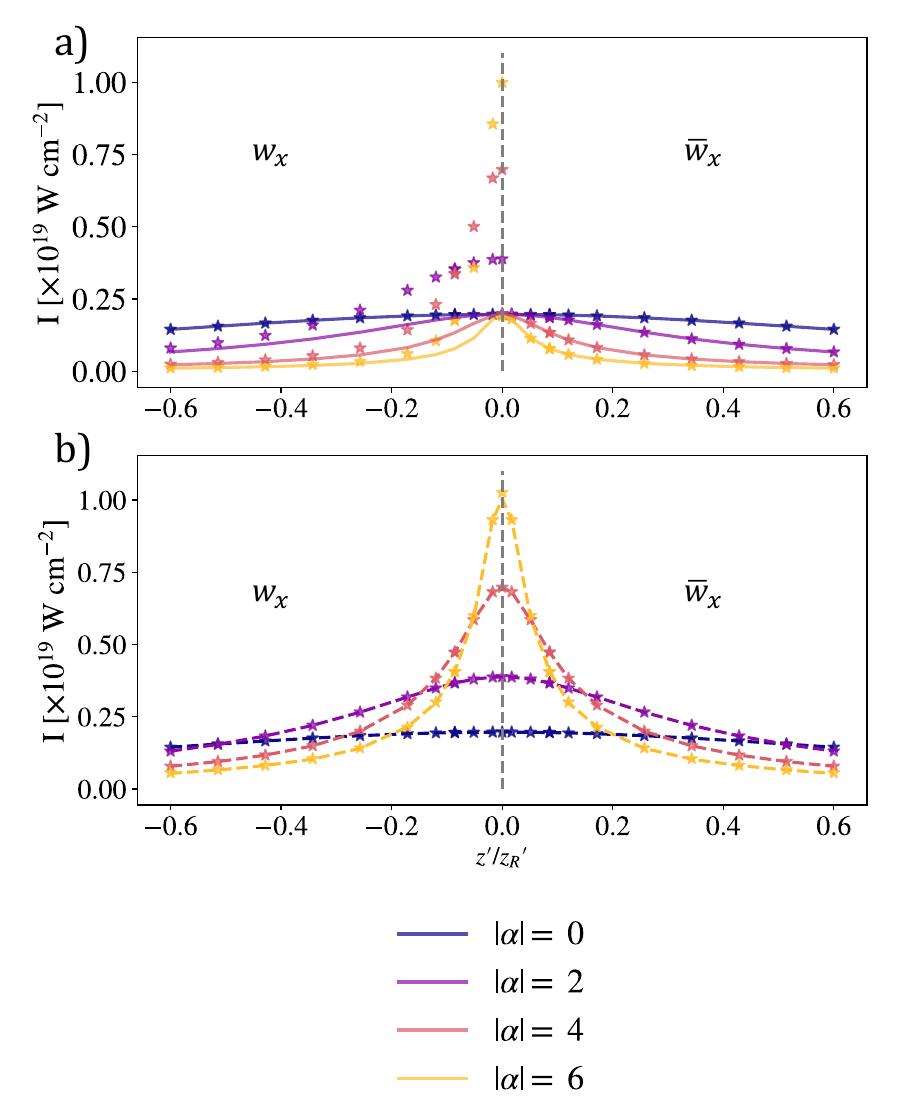}
    \caption{Comparison of the heuristic EFV method (stars) given by Eq.\,\ref{Eq:EFVPI} using both the instantaneous spot size $w_x(z)$ (left side) and the fluence-based spot size $\overline{w}_x(z)$ (right side). The EFV methods are plotted against the correct peak axial intensities for (a) SSTF pulses (solid lines); and (b) SE pulses (dashed lines).}
    \label{fig:intensityefv}
\end{figure}

\subsection{Heuristic methods for calculating the intensity of SSTF pulses}\label{Sec:EFV}
In the section above we showed that the region of high intensity of a SSTF pulse is no more localised than that of the spatially equivalent pulse. This may seem to contradict the intuitive statement that shorter laser pulses have higher peak intensities. Here we use a simple model to show that such logic fails for pulses with spatiotemporal couplings.

The intensity of a bi-Gaussian laser pulse may be written in the form \cite{siegman}: 

\begin{equation}
    I(x,y,z,\tau) \propto \frac{1}{w_x(z)w_y(z)}\exp\left[- \frac{x^2}{w_x(z)^2} - \frac{y^2}{w_x(z)^2}\right] \exp \left[- 2 \ln 2 \left(\frac{\tau}{\tau_\mathrm{p}(z)}\right)^2\right].
\end{equation}
For a pulse of this form, the loci of constant intensity are ellipses aligned to the $x$- and $y$-axes. In terms of the energy $\mathcal{E}$ of the pulse, the peak on-axis intensity can be written as,

\begin{equation}
    I^\mathrm{EFV}_{\textrm{peak}}(z) = 4\sqrt{\frac{\ln{2}}{\pi^3}}\frac{\mathcal{E}}{w_x(z) w_y(z) \tau_{\textrm{p}}(z)}.\label{Eq:EFVPI}
\end{equation}
Eq.\,\ref{Eq:EFVPI} is commonly used to calculate the longitudinal variation of the on-axis intensity for a bi-Gaussian laser pulse, which we shall call the ellipsoidal focal volume (EFV) method.

The left side of Fig.~\ref{fig:intensityefv} compares the longitudinal variation of the peak axial intensities of the SSTF and SE pulses calculated directly from the fields with those calculated from Eq.\,\ref{Eq:EFVPI}, where $w_x(z)$ and $w_y(z)$ were determined from the numerically-calculated intensity distributions, as described in \S \ref{Sec:Numerical_treatment}. It can be seen from the left side of Fig.~\ref{fig:intensityefv}(a) that the EFV method overestimates in the degree of axial localisation of the high intensity region of the SSTF pulse, and (incorrectly) suggests that the SSTF pulse is more localised than the SE pulse. In contrast, as evident from the left-side of Fig.~\ref{fig:intensityefv}(b), the EFV method \emph{does} correctly calculate the longitudinal intensity variation of the SE pulse.

\begin{figure}[!tb]
    \centering
    \includegraphics[width=.8\linewidth]{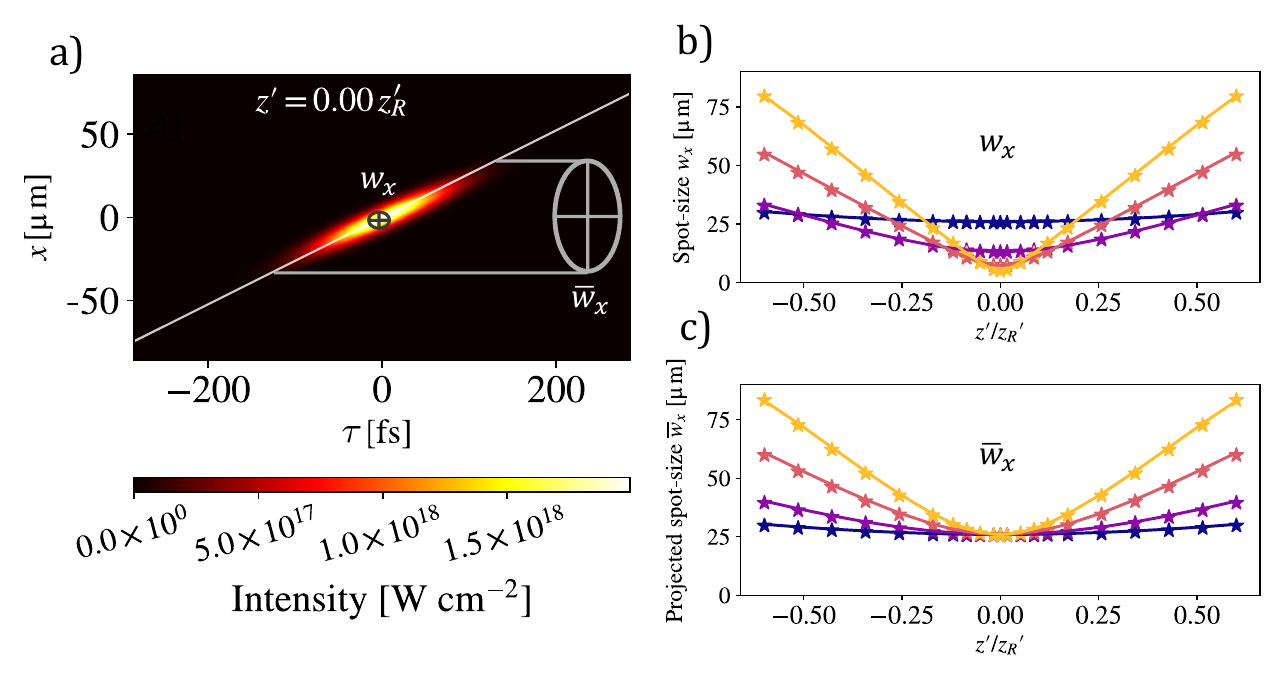}
    \caption{Illustration of the reason why EFV methods might fail to calculate correctly the intensity of a SSTF pulse. (a) Shows the spatio-temporal intensity profile at $z'=0$ of an SSTF pulse with $\alpha=-3$. The solid white line shows the PFT predicted by Eq.\,\ref{Eq:PFTevo}. The black lines superimposed on the spatio-temporal intensity distribution illustrate the estimated focal volume, assuming the instantaneous spot size $w_x(z)$ and pulse duration $\tau_\mathrm{p}(z)$; whereas the grey lines illustrate the estimated focal volume assuming the fluence-based spot size $\overline{w}_x(z)$ obtained from the fluence. Comparison of (b) the instantaneous spot size $w_x(z)$ calculated directly from the intensity (lines) with the analytical expression given by Eq.\,\ref{Eq:InstSpot} (stars) and (c) the fluence-based spot size $\overline{w}_x(z)$ calculated directly from the fluence (lines) with the analytical expression given by Eq.\,\ref{Eq:FluenceSpot} (stars). In (b) and (c) the colour scheme is the same as in Figs.~\ref{fig:outputlinearparameters}--\ref{fig:intensityefv}.}
    \label{fig:focalvolume}
\end{figure}

The reasons for this failure are evident from Fig.~\ref{fig:focalvolume}. This shows that, when used with a spot size $w_x(z)$, the EFV method significantly overestimates the peak intensity of a pulse with PFT since it underestimates the focal volume of the pulse. However, if the \emph{fluence-based} spot size $\overline{w}_x(z)$ is used instead, then the EFV method correctly calculates the peak axial intensity for both SSTF and SE pulses, as can be seen from the right side of Fig.~\ref{fig:intensityefv}. Here, the fluence-based spot size $\overline{w}_x(z)$ is calculated from the second moment of the transverse fluence distribution, $F^\mathrm{out}(x,y,z) = \int_{-\infty}^\infty \mathrm{d} \tau \, I^\mathrm{out}(x,y,z,\tau)$, rather than from the transverse intensity profile at the peak of the pulse.

The analytic model gives the instantaneous spot size as,
\begin{equation}
    w_x'(z) = \frac{w'(0)}{\sqrt{1 + \frac{4 \alpha^2 \Gamma}{w_0^2}}} \sqrt{1 + \left(\frac{z'}{z'_\mathrm{R,eff}}\right)^2},\label{Eq:InstSpot}
\end{equation}
and the fluence-based spot size as,
\begin{equation}
    \overline{w}_x(z) = w'(0) \sqrt{1 + \left(\frac{z'}{\sqrt{z'_\mathrm{R}z'_\mathrm{R,eff}}}\right)^2}.\label{Eq:FluenceSpot}
\end{equation}
The spot size of each Gaussian beamlet after the lens is given by,

\begin{align}
    w'(z) &= w'(0)\sqrt{1 + \left(\frac{z'}{z'_\mathrm{R}} \right)^2},
\end{align}
where $w'(0) = \sqrt{\lambda_0 z'_\mathrm{R}/\pi}$. For beams without PFT, such as the SE pulse, the instantaneous and fluence-based spot sizes are the same, such that $\overline{w}_x(z) = w_x(z)$, as evident in Fig.~\ref{fig:intensityefv}(b). In this case, the instantaneous and fluence-based spot sizes are both equal to the instantaneous spot size of the SSTF pulse.

For SSTF beams, each beamlet is focused to a waist at $z = f_0$ with a spot size in the $x$-direction of $w'(0)$. The SSTF pulse is formed from a superposition of these beamlets, and, as such, the superposition is expected to be constrained to a transverse size (in the $x$-direction) that is set by $w'(0)$. It is clear from Eq.\,\ref{Eq:FluenceSpot} that the the focal spot size of the \emph{fluence} distribution $\overline{w}_x(0)$ is equal to $w'(0)$. This is independent of $\alpha$ as seen in Fig.\ \ref{fig:focalvolume}(c). In the $x$-direction the spot size of the fluence distribution increases with distance from the focus with an effective Rayleigh range $\sqrt{z'_\mathrm{R}z'_\mathrm{R, eff}}$. Although the beamlets overlap in the focal plane, and have a transverse size (in the $x$-direction) set by $w'(0)$, for an SSTF pulse the beamlets associated with different frequencies propagate at different angles to the system axis. As a consequence, the resulting spatio-temporal distribution of the fields exhibits a pulse-front tilt, and hence the transverse intensity profile at the peak of the pulse has a spot size $w_x(0)$ that is smaller than that of each beamlet, i.e.\ $w_x(0) < w'(0)$. Increasing the magnitude of the spatial chirp (i.e. increasing $|\alpha|$) will increase the magnitude of the PFT, and hence will decrease the instantaneous spot-size $w_x(0)$. This behaviour is evident from Eq.\,\ref{Eq:InstSpot} and Fig.\ \ref{fig:focalvolume}(b).

We note also that at large distances from the focus the PFT of the SSTF pulse becomes small, so that the instantaneous and projected spot sizes become equal. This is seen in Fig.\ \ref{fig:focalvolume}(b) and (c) and is consistent with Eqs.\,\ref{Eq:InstSpot} and \ref{Eq:FluenceSpot}.

\section{Conclusion}
In this work we have attempted to address some confusion in the literature about the degree, and causes, of localisation of the high-intensity region of a SSTF pulse. To this end we used numerical and analytical methods to investigate the degree of localisation of the spatio-spectral and spatio-temporal amplitudes and intensities of a SSTF pulse formed from a non-temporally-chirped, bi-Gaussian input pulse. The analytical model was found to be in excellent agreement with the numerical simulations, and also yielded expressions for the three-dimensional spatio-temporal intensity profile and the on-axis bandwidth, pulse duration, and pulse-front tilt of the SSTF pulse. 

In order to compare the behaviour of SSTF and non-SSTF pulses, we proposed a general definition for the non-SSTF pulse with spatially-equivalent focusing. Our calculations show that the duration of a SSTF pulse increases with distance away from the focus, as expected. However, despite this, the degree of localisation of the high-intensity region of a SSTF pulse was found, for the bi-Gaussian input pulses we considered, to be no better than that of the spatially-equivalent (SE) pulse. As such the SSTF method offers no advantage in constraining the region of high intensity over focusing a non-SSTF pulse with equivalent focusing, in agreement with the original literature on this subject \cite{Durst:06,j.optcom.2007.05.071}.

We found that a simple, heuristic method can correctly calculate the longitudinal intensity variation if the spot size is calculated from the fluence of the pulse, but that it overestimates the degree of intensity localisation if the spot size is calculated from the transverse intensity profile at the peak of the pulse. This difference in behaviour is caused by the PFT exhibited by SSTF pulses in the focal region. 

In this work we showed that SSTF does not provide tighter localization of the region of high pulse intensity than a non-SSTF pulse with spatially-equivalent focusing, and hence offers no advantages for purely intensity-dependent processes. However, SSTF \emph{does} provide several advantages in other contexts. For example, for purely fluence-dependent processes, SSTF allows the effective Rayleigh range (or, equivalently, the depth of field) to be reduced whilst keeping the fluence-based spot size constant. Further, as discussed by Durfee et al. \cite{OE.20.014244}, SSTF increases the pulse duration, and reduces the peak intensity, in the regions surrounding the focus, which can reduce unwanted nonlinear effects such as self-phase modulation or self-focusing.

We emphasize that our results were obtained for the special case of non-temporally-chirped, bi-Gaussian input pulses. The numerical method and techniques for benchmarking pulses against their conventional counterparts that we outline could readily be extended to other spatial and spectral profiles, including structured spatiotemporal optical wavepackets with additional spatiotemporal couplings \cite{Kondakci2019,PhysRevA.103.033512,Jolly:23,Liu2024}, and the analytical approach could be extended to the case of temporally-chirped input pulses. We would expect, however, that the main conclusions of the present paper would broadly apply for more general cases.

\section{Backmatter}

\begin{backmatter}
\bmsection{Funding}
This work was supported by the UK Engineering and Physical Sciences Research Council (EPSRC) (Grant No.\ EP/V006797/1 and EP/R513295/1), the UK Science and Technologies Facilities Council (Grant No.\ ST/V001655/1). For the purpose of Open Access, the author has applied a CC BY public copyright licence to any Author Accepted Manuscript version arising from this submission.

\bmsection{Disclosures}
The authors declare that there are no conflicts of interest related to this article.

\bmsection{Data Availability Statement}
Data underlying the results presented in this paper may be obtained from the authors upon reasonable request.

\bmsection{Supplemental document}
See \underline{Supplement 1} for supporting content

\end{backmatter}


\bibliography{ref}

\end{document}